\documentclass[aps,amssymb,prl,twocolumn,showpacs]{revtex4}
\usepackage{graphicx}
\usepackage{epsfig,amsmath}
\pdfoutput=1
\begin{document}

\title{Selection-rule blockade and rectification in quantum heat transport}
\author{Teemu Ojanen}
\email[Correspondence to ]{ojanen@physik.fu-berlin.de}
\affiliation{ Institut f\"ur Theoretische Physik, Freie Universit\"at Berlin,
Arnimallee 14, 14195 Berlin, Germany}

\date{\today}
\begin{abstract}
We introduce a new thermal transport phenomenon, a unidirectional selection-rule blockade, and show how it produces unprecedented rectification of bosonic heat flow through molecular or mesoscopic quantum systems. Rectification arises from the quantization of energy levels of the conduction element and selection rules of reservoir coupling operators. The simplest system exhibiting the selection-rule blockade is an appropriately coupled three-level system, providing a candidate for a high-performance heat diode. We present an analytical treatment of the transport problem and discuss how the phenomenon generalizes to multilevel systems.
\end{abstract}
 \bigskip

\maketitle

Heat conduction in nanoscale structures has become an active field of research enjoying constantly increasing attention. Electronic properties have been studied in great detail in the last three decades and are better understood than thermal properties. The main reason for this has been experimental challenges to measure and control thermal properties accurately. Recently the field has seen breakthroughs such as measurements of quantized heat transport \cite{schwab,meschke} and realization of hybrid structures probing atomic-level heat transport properties \cite{carey}.

Experimental developments have increased theoretical interest to explore fundamental limits of thermal phenomena and devices based on their applications. Thermal rectification, i.e. asymmetry of heat current when the temperature bias is inverted, has been actively studied in this context. Rectification has been observed so far in two experiments \cite{chang,scheibner} and has a significant application potential. There exists various theoretical proposals of how to realize rectification in phononic \cite{terraneo,hu}, photonic \cite{ruokola} and other hybrid structures \cite{segal}. However, in most proposals rectification is modest and limits of performance are unknown. The purpose of this paper is to introduce a novel heat transport phenomenon, an asymmetric selection-rule blockade, enabling an unparalleled rectification performance. The phenomenon provides an example of an intricate interplay of quantum-mechanical and thermal properties in nanoscale structures.

The simplest system exhibiting the selection-rule blockade is a three-level system with a strongly uneven energy-level separations. A crucial ingredient is that the two baths are coupled to the system so that one can only induce transitions between the close-lying states and the other can effectively induce only the two larger transitions. The baths can exchange energy only when the bath coupling far-apart states has sufficiently high temperature to create excitations. In suitable temperature regime the forward biasing leads to sequential heat flow while the reverse bias current, due to higher-order processes, is very strongly suppressed. This \emph{one-directional} suppression of heat flow is the defining property of the selection-rule blockade.

The paper is organized as follows. First we introduce the three-level model and show analytically how rectification arises from the selection rules. Then we estimate the magnitude of the leakage current that sets the limits of performance of rectification and discuss how the selection rules can be realized without special symmetries. We conclude by outlining how the selection-rule blockade generalizes to multilevel systems and summarize our results.

\begin{figure}[h]
\begin{center}
\includegraphics[width=0.55\columnwidth]{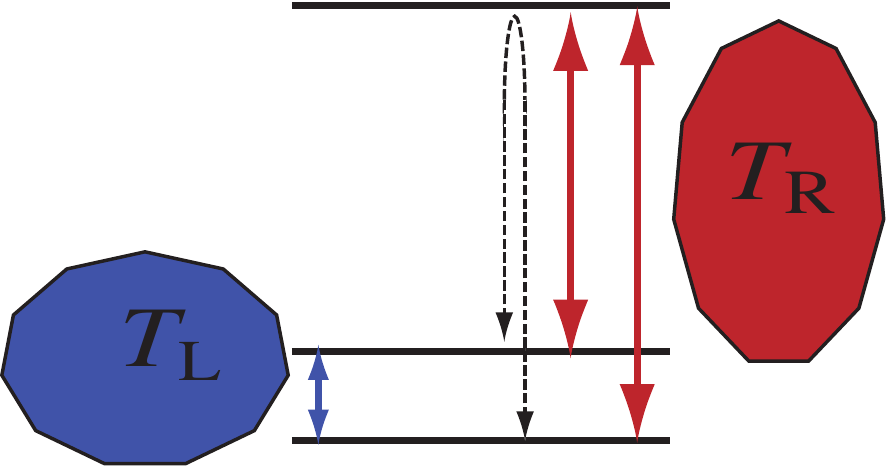}
\caption{Studied model consists of two heat baths at different temperatures coupled through a three-level system. The blue arrow represents transitions induced by the left reservoir and the red arrows correspond to Golden-Rule transitions induced by the right reservoir. The dashed arrow represents higher-order transitions via the virtual state by the right reservoir.}
\label{scheme}
\end{center}
\end{figure}

In the following we consider thermal conduction in the setup depicted in Fig.~\ref{scheme}. The system consists of three parts, two reservoirs and a three-level system that mediates heat. We call a configuration forward biased if $T_{\rm R}>T_{\rm L}$ and reverse biased in the opposite case. Intuitively one expects that if temperatures $T_{\rm L/R}$ are much smaller than the allowed transition energies, heat flow is effectively blocked. This is confirmed below as we show how the intrinsic asymmetry in the strength of the blockade gives rise to strong rectification. The Hamiltonian of the system is
\begin{align}\label{total}
H=H_{{\rm L}}+H_{{\rm R}}+H_{{\rm C}}+H_{{\rm LC}}+H_{{\rm RC}}
\end{align}
where $H_{{\rm L/R}}$ characterize the reservoirs and the central Hamiltonian is
\begin{align}\label{cent}
H_{{\rm C}}=\epsilon_0|0\rangle\langle0|+\epsilon_1|1\rangle\langle1|+\epsilon_2|2\rangle\langle2|.
\end{align}
The level separation  of the central part $E_1=\epsilon_1-\epsilon_0$, $E_2=\epsilon_2-\epsilon_1$ is assumed to be unequal $E_1/E_2\ll 1$.
The reservoirs couple to the central system in a special manner
\begin{align}\label{coup}
&H_{{\rm LC}}=X_{\rm L}\left( |0\rangle\langle1|+|1\rangle\langle0| \right)\nonumber\\
&H_{{\rm RC}}=X_{\rm R}\left( |1\rangle\langle2|+|2\rangle\langle1|+|0\rangle\langle2|+|2\rangle\langle0| \right),
\end{align}
where $X_{\rm L/R}$ are Hermitian operators in the reservoir Hilbert spaces.
The specific form of the couplings imply that the left reservoir couples \emph{only} the groundstate and the first excited state in contrast to the right reservoir which couple all the energy levels. Applying the Fermi Golden Rule one can calculate the transition rates between the energy levels
\begin{align}\label{gamma1}
&\Gamma_{1\to 0}={S_{X_{\rm L}}(\omega_1)}/{\hbar^2} ,\qquad
\Gamma_{2\to 1}={S_{X_{\rm R}}(\omega_2)}/{\hbar^2},\nonumber \\
&\Gamma_{2\to 0}={S_{X_{\rm R}}(\omega_1+\omega_2)}/{\hbar^2},
\end{align}
where $S_{X_{\rm L/R}}(\omega)=\int_{-\infty}^{\infty}dt e^{i\omega\, t}\langle X_{\rm L/R}(t)X_{\rm L/R}(0)\rangle $ and $\omega_{1/2}=E_{1/2}/\hbar$. The correlation functions are evaluated in the absence of the couplings at the respective reservoir temperatures. Information about $H_{{\rm L}}$ and $H_{{\rm R}}$ as well as information about the structure of the reservoirs is encoded in the noise $S_{X_{\rm L/R}}(\omega)$ so our discussion is independent of the precise nature of the reservoirs so far. It should be noted that even though the right reservoir can induce direct transitions between the ground state and the first excited state it happens only in higher-orders. In the lowest order the allowed transitions are separate in the different reservoirs. The corresponding inverse transitions are obtained by inverting the sign of the frequency argument, for example
$\Gamma_{0\to 1}=S_{X_{\rm L}}(-\omega_1)/\hbar^2$. Noise is related to the retarded $X$-correlation function and the Bose-Einstein distribution $n(\omega)$ through the Fluctuation-Dissipation Theorem
\begin{align}\label{FD}
S_{X_{\rm L/R}}(\omega)=A_{\rm L/R}(\omega)(1+n_{\rm L/R}(\omega)),
\end{align}
where $A_{\rm L/R}(\omega)=-2\,{\rm Im} \langle X_{\rm L/R}X_{\rm L/R}\rangle^r(\omega)$ are spectral densities of the reservoirs defined through the Fourier transform of the retarded function
$\langle X_{\rm L/R}(t)X_{\rm L/R}(0)\rangle^{r}=-i\theta(t)\langle[X_{\rm L/R}(t),X_{\rm L/R}(0)]\rangle$ \cite{bruus}.
Once the transition rates are known, the steady-state occupation probabilities follow from the detailed-balance equations
\begin{align}\label{db}
-P(i)\sum_{j\neq i}\Gamma_{i\to j}+\sum_{j\neq i}P(j)\Gamma_{j\to i}=0.
\end{align}
Employing the rate-equation formulation \cite{bruus}, heat current can be calculated, say, between the left reservoir and the central system yielding
\begin{align}\label{cur1}
J=E_1\left[ P(1)\Gamma_{1\to 0}-P(0)\Gamma_{0\to 1} \right].
\end{align}
The probabilities required to evaluate (\ref{cur1}) can be solved in terms of the transition rates as
\begin{align}\label{prob}
&P(0)=\left({\Gamma_{1\to 0}\Gamma_{2\to 0}+\Gamma_{1\to 2}\Gamma_{2\to 0}+\Gamma_{1\to 0}\Gamma_{2\to 1}}\right)/{C}\nonumber \\
&P(1)=\left({\Gamma_{0\to 1}\Gamma_{2\to 0}+\Gamma_{0\to 1}\Gamma_{2\to 1}+\Gamma_{0\to 2}\Gamma_{2\to 1}}\right)/{C},
\end{align}
with
\begin{align}\label{prob}
&C=\Gamma_{10}\Gamma_{20}+\Gamma_{12}\Gamma_{20}+\Gamma_{10}\Gamma_{21}+\Gamma_{01}\left(\Gamma_{20}+\Gamma_{12}+\Gamma_{21}\right)\nonumber \\
&+\Gamma_{02}\left(\Gamma_{21}
+\Gamma_{10}+\Gamma_{12}\right),
\end{align}
where we have introduced notation $\Gamma_{i\to j}=\Gamma_{ij}$.
Inserting the probabilities and the transition rates in Eq.~(\ref{cur1}) we obtain an explicit expression for the current
\begin{align}\label{cur2}
&J=E_1\frac{\Gamma_{01}\Gamma_{02}\Gamma_{21}}{C}(e^{\beta_LE_1}-e^{\beta_RE_1}).
\end{align}
Result (\ref{cur2}) indicate that the thermal window $(e^{\beta_LE_1}-e^{\beta_RE_1})$ determined by the reservoir temperatures and the smaller energy separation $E_1$ plays an important role in the transport process.
\begin{figure}[h]
\begin{center}
\includegraphics[width=0.75\columnwidth]{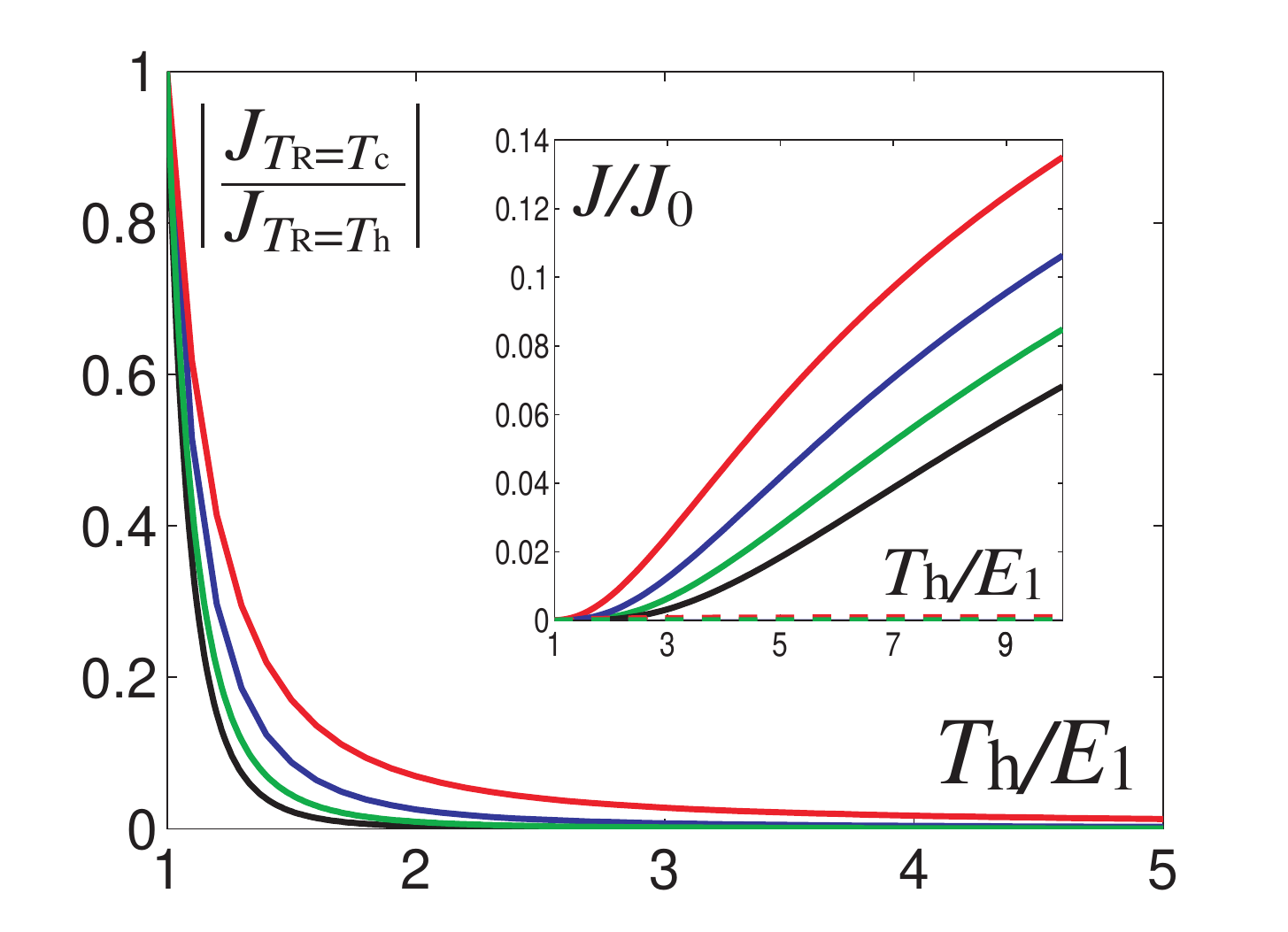}
\caption{Rectification efficiency as a function of the higher operation temperature. The different curves correspond to values $T_{\rm c}=E_1$, and $E_2/E_1=5$ (red), $E_2/E_1=7$ (blue), $E_2/E_1=9$, (green) and $E_2/E_1=11$ (black). Inset illustrates the corresponding forward bias (solid lines) and reverse bias currents (dashed lines, almost zero) in the units of $J_0=E_1A_0/\hbar^2$.}
\label{plotti}
\end{center}
\end{figure}
In the limit $k_{\rm B}T_{\rm R}\ll E_2$ current (\ref{cur2}) takes a simple form
\begin{align}\label{cur3}
J_{T_{\rm R}\ll E_2}=E_1\frac{A_{\rm R}(\omega_2)}{1+\frac{A_{\rm R}(\omega_2)}{A_{\rm R}(\omega_1+\omega_2)}}
\frac{e^{-\beta_R(E_1+E_2)}(e^{\beta_LE_1}-e^{\beta_RE_1})}{(1+e^{\beta_LE_1})}.
\end{align}
Result (\ref{cur3}) shows that heat current is exponentially suppressed in the considered limit. This is an indication that the system acts as a high-performance thermal rectifier when the lower operation temperature is much smaller than $E_2$ while the higher temperature is comparable to it. To make rectification more explicit, consider a situation where the hot reservoir is at temperature $k_{\rm B}T_{\rm h}\sim E_2$ and the cold one is at $k_{\rm B}T_{\rm c}\sim E_1$. Furthermore, to simplify expressions, let us assume that the effective couplings are of the same order $A_{\rm L}(\omega_1)\approx A_{\rm R}(\omega_2)\approx A_{\rm R}(\omega_1+\omega_2)=A_0$. Starting from the expression (\ref{cur2}) one can estimate that the fraction of the bias-inverted heat currents is
\begin{align}\label{rec}
\left|\frac{J_{T_{\rm R}=T_{\rm c}}}{J_{T_{\rm R}=T_{\rm h}}}\right|=c\,e^{-E_2/E_1}\sim c\,e^{-T_{\rm h}/T_{\rm c}},
\end{align}
where $c$ is a numerical factor of the order of unity. Thus Eq.~(\ref{rec}) shows that in the studied temperature regime the fraction of forward and reverse bias currents is \emph {exponentially small} in $T_{\rm h}/T_{\rm c}$. Results (\ref{cur2})-(\ref{rec}) illustrate the essence of the selection rule blockade and confirm quantitatively the intuitive physical picture. More comprehensive picture of transport properties can be obtained by plotting current (\ref{cur2}) and the rectification ratio at different forward and reverse bias values, see Fig.~\ref{plotti}. Reverse bias current is rapidly suppressed as the ratio $E_2/E_1$ is increased and \emph{effectively vanishes} in the temperature window $k_{\rm B}T_{\rm c}\leq E_1$, $k_{\rm B}T_{\rm h}>3E_1$ for $E_2/E_1>3$. Importantly, forward bias current is finite and increasing in the region of optimal rectification.

In the derivation of Eq.~(\ref{cur2}) we only took into account the Golden-Rule transitions. Deep in the blockaded regime $k_{\rm B}T_{\rm R}\ll E_2$ the lowest-order current vanishes exponentially as shown in Eq.~(\ref{cur3}) and one is naturally lead to consider higher-order transitions. The situation is similar to electronic transport in the Coulomb blockade model where charging effects suppress conductance (with the difference that the selection-rule blockade is one directional). At low temperatures and biases electric current through a small island vanishes exponentially in the lowest order in the perturbation theory. However, in the next order one obtains cotunneling processes where electrons are transferred via virtual intermediate states partly lifting the blockade \cite{averin}. In our model the leading higher-order processes contributing to the reverse leakage current in the blockaded regime correspond to the right reservoir-induced transitions via the highest-energy state as depicted in Fig.~\ref{scheme}. These rates can be calculated, for example, by the standard $T$-matrix expansion to the second order \cite{bruus}. Let us assume that the right reservoir consists of noninteracting bosonic modes and that the coupling operator is of the form $X_{\rm R}=\sum_{j\in \rm{R}}c_j(b_j+b_j^{\dagger})$ where $b_j,\,b_j^{\dagger}$ are canonical bosonic operators of the mode $\omega_j$. In the low-temperature limit $T_{\rm R}\to 0$ the relevant second-order process contributing to reverse bias current is the decay of the first excited state to the ground state by emitting two bosons to the right reservoir. This process gives rise to an additional rate
\begin{align}\label{second1}
&\Gamma^{(2)}_{1\to 0}=\frac{2\pi}{\hbar^2}\int_0^{\omega_1} d\omega A_{\rm{R}}(\omega)A_{\rm{R}}(\omega_1-\omega)\times\nonumber\\
&\left|\frac{1}{E_2-\hbar\omega}+\frac{1}{E_2-E_1+\hbar\omega} \right|^2\nonumber\\ &\approx\frac{8\pi}{\hbar^2E_2^2}\int_0^{\omega_1} d\omega A_{\rm{R}}(\omega)A_{\rm{R}}(\omega_1-\omega).
\end{align}
The rate is proportional to a typical second-order energy denominator $\sim 1/E_2^2$ suppressing the transition. Assuming that the spectral density has a power-law form $A_{\rm{R}}(\omega)\propto\omega^d$ $(d=1$ for an ohmic bath), the leading contribution becomes
\begin{align}\label{second2}
&\Gamma^{(2)}_{1\to 0}=\frac{8\pi}{1+d}\frac{(\hbar\Gamma_{2\to 0\, |T_{\rm R}=0 })^2}{E_2^2}\left(\frac{E_1}{E_2}\right)^{2d}\omega_1.
\end{align}
The second and the third factor are much smaller than unity so (\ref{second2}) is small compared to the level separation $\omega_1$. However, the rate is only algebraically suppressed by $(E_1/E_2)^{2d}$. At finite temperature $T_{\rm R}\sim E_1$ the relevant power law exhibits a crossover to $(k_{\rm{B}}T_R/E_2)^{2d}$. Now we can estimate the leakage current under the blockade by
\begin{align}\label{cur4}
J_{\rm r}=-E_1 P(1)\Gamma_{1\to 0}^{(2)}=-E_1\frac{\Gamma_{1\to 0}^{(2)}}{1+e^{\beta_{\rm L}E_1}+\frac{\Gamma_{1\to 0}^{(2)}}{\Gamma_{1\to 0}}}.
\end{align}
Result (\ref{cur4}) shows that the  true rectification efficiency is algebraically, not exponentially, suppressed. It is clear that the leakage current remain finite as long as there is thermal coupling between the two reservoirs and that a good rectifier is characterized by its ability to suppress the parasitic reverse bias processes. Nevertheless, current (\ref{cur4}) can be made arbitrarily small by decreasing  max$\{E_1/E_2,k_{\rm B}T_{\rm R}/E_2\}$.



\begin{figure}[h]
\begin{center}
\includegraphics[width=0.87\columnwidth]{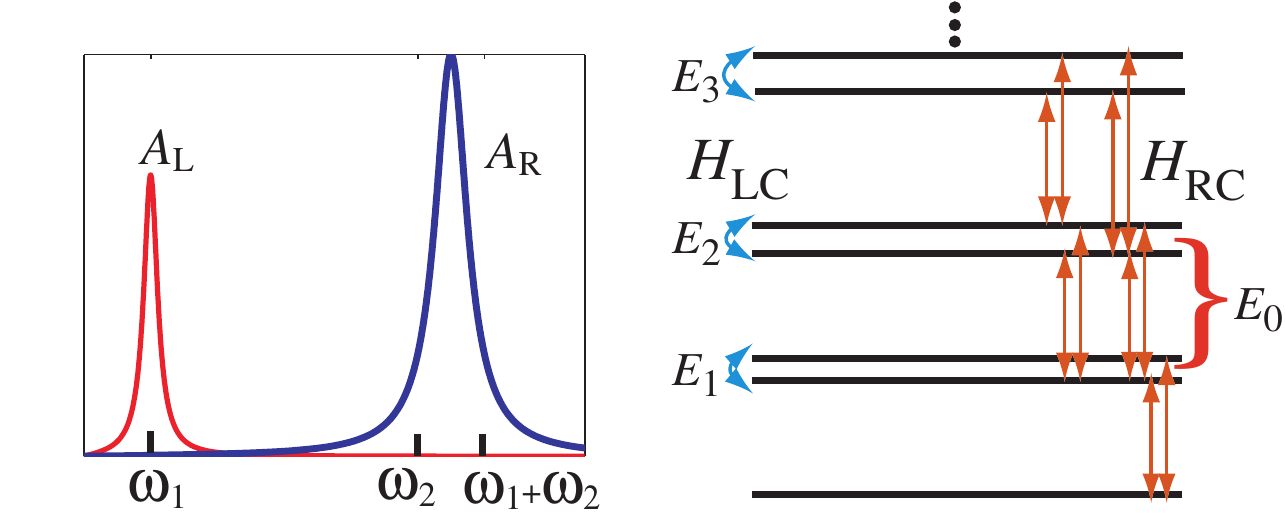}
\caption{Left: Spectral density of the left (right) reservoir is completely suppressed at frequencies $\omega\gg\omega_1$ ($\omega\ll\omega_2$), effectively imposing the desired selections rules. Right: Low-lying spectrum of the Jaynes-Cummings model and the allowed transitions induced by the left (blue arrows) and the right (red arrows) reservoir.}
\label{vika}
\end{center}
\end{figure}

Usually selection rules are associated with symmetries of the Hamiltonian leading to vanishing matrix elements of the perturbation operator between unperturbed states. In applications finding a candidate system with suitable symmetries and couplings is a nontrivial task. However, since the transitions of the different reservoirs are well separated in energy, there exists an alternative route to obtain the selection rules without invoking symmetries of the Hamiltonian.
Golden-Rule rates consists of matrix elements of perturbation operators and a summation over a relevant phase space. Both factors can effectively impose selection rules in the system. Typically phase-space selection rules arise from the fact that the reservoir spectral density is nonvanishing only in a restricted frequency window outside which the reservoir cannot induce transitions. In solid-state applications suitable reservoirs can be realized by dissipative vibrational modes of phononic or photonic nature which filter out frequencies far from the resonance. Assuming that the center element couples linearly to the modes, the reservoir parts of the couplings (\ref{coup}) take the form $X_{\rm L/R}=c_{\rm L/R}(b^\dagger_{\rm L/R}+b_{\rm L/R})$, where $b^\dagger_{\rm L/R},\,b_{\rm L/R}$ are the creation and annihilation operators of the reservoir modes.  The retarded Green's function of a vibrational mode is $G^r(\omega)=((g_0^r)^{-1}+\Sigma^r(\omega))^{-1}$, where $(g_0^r)^{-1}=(\omega^2-\omega_0^2)/2\omega_0$ is the inverse free Green's function, $\omega_0$ is the frequency of the mode and $\Sigma^r(\omega)$ is the retarded self-energy due to dissipation.  Dissipation can be realized by coupling the mode to a bosonic bath, in which case the spectral densities $A_{\rm L/R}(\omega)=-2g_{\rm L/R}^2{\rm Im}\, G^r(\omega)$ of the reservoirs can be solved exactly \cite{ojanen2} yielding
\begin{align}\label{spect}
A_{\rm L/R}(\omega)=\frac{c_{\rm L/R}^2(2\hbar\omega_{\rm L/R})^2A^{\rm B}_{\rm L/R}(\omega)}{\hbar^4(\omega^2-\omega_{\rm L/R}^2)^2+(A^{\rm B}_{\rm L/R}(\omega)\omega_{\rm L/R})^2},
\end{align}
where $\omega_{\rm L/R}$ are the resonant frequencies of the reservoirs and $A^{\rm B}_{\rm L/R}(\omega)$ are the spectral densities responsible for dissipation of the reservoir modes. As illustrated in Fig.~\ref{vika} (left), if the left reservoir has a resonance frequency at $\omega_L=\omega_1$ and the right reservoir at $\omega_R=\omega_2+\omega_1/2$ with respective widths \footnote{Widths of the resonances of $A_{\rm L/R}(\omega)$ are determined by the strength of dissipation through $A^{\rm B}_{\rm L/R}(\omega)$} that are much smaller than $\omega_2$, the spectral densities have no overlap. Thus the reservoirs can effectively induce transitions only as indicated in Fig.~\ref{scheme}. If the reservoirs are electromagnetic in nature the desired spectral densities (\ref{spect}) are achieved by coupling the center element to dissipative LC-circuits whose electromagnetic fluctuations are restricted to a narrow band around the resonance \cite{niskanen}. In phononic systems the same effect could be realized by coupling large baths to the central element through small vibrating bridges.

The selection-rule blockade is a not restricted to the studied three-level system. Basic requirements are that level separations of a system can be divided to small and large intervals coupled to separate baths, and that combined effect of two large transitions corresponds to a small one as in the tree-level model. A prominent example fulfilling the requirements is a two-level system coupled to a harmonic oscillator described by a resonant Jaynes-Cummings- type Hamiltonian \cite{breuer}
\begin{align}\label{JC}
H_{\rm JC}=\hbar\Omega\left(a^\dagger a+\frac{1}{2}\right)+\frac{\hbar\Omega}{2}\sigma_z+\frac{\hbar g}{2}(a^\dagger+a)\sigma_x,
\end{align}
where the interaction is assumed to be small $g\ll\Omega$. Excited states of the Hamiltonian (\ref{JC}) consists of doublets $|E_n\rangle=|n,\downarrow\rangle\pm|n-1,\uparrow\rangle $ where the energy separation of centers of adjacent doublets is approximately $E_0=\hbar\Omega$ and the separation within a doublet is $E_{n}=\sqrt{n}\hbar g$, see Fig.~\ref{vika} (right). Since for the low-lying doublets energy splittings are small $E_n\ll E_0$, the selection-rule blockade can be established if the left reservoir couples levels only within doublets and the right reservoir couples states in different doublet. This requirement is satisfied in the lowest order by $H_{{\rm RC}}=X_{\rm R}(a^\dagger+a)$ and $H_{{\rm LC}}=X_{\rm L}\sigma_z$, yielding immediately the desired selection rules. Analogous to the three-level model, if the temperatures of the reservoirs are $k_{\rm B}T_{\rm L}\sim\hbar g$, $k_{\rm B}T_{\rm R}\sim\hbar\Omega$ heat flow is efficient, while in the reverse biased case $k_{\rm B}T_{\rm L}\sim\hbar\Omega$, $k_{\rm B}T_{\rm R}\sim\hbar g$ excitations induced by the right reservoir are forbidden and there exists only a weak leakage current due to higher-order processes.

In conclusion, we introduced a fundamental heat transport phenomenon, the selection-rule blockade, based on the quantum nature of the central system and selection rules of the bath coupling operators. The phenomenon enables a high rectification efficiency in molecular and mesoscopic structures making it promising for future device applications. The simplest system exhibiting an asymmetric blockade is a three-level model which we analyzed in detail. We proposed a scheme to impose required selection rules without special symmetry properties of the Hamiltonian and discussed how the selection-rule blockade can be generalized to multilevel systems.

\end{document}